\documentclass[pra,twocolumn,floatfix,superscriptaddress,showpacs]{revtex4}
\usepackage{graphicx}
\usepackage{amssymb}
\usepackage{amsmath}

\newcommand{\ssm}{\scriptscriptstyle\rm}
\newcommand{\ssmR}{{\ssm R}}
\newcommand{\ssmL}{{\ssm L}}
\newcommand\pdag{{\vphantom \dagger}}

\begin{document}

\title{Coherent pumping of a Mott insulator: Fermi golden rule versus Rabi
oscillations}

\author{F. Hassler}
\affiliation{Institute for Theoretical Physics, ETH Zurich, 8093
Zurich, Switzerland}
\author{S.~D. Huber}
\affiliation{Department of Condensed Matter Physics, The Weizmann
Institute of Science, Rehovot, 76100, Israel}
\affiliation{Institute for Theoretical Physics, ETH Zurich, 8093
Zurich, Switzerland}

\date{\today}

\begin{abstract}
Cold atoms provide a unique arena to study many-body systems far from
equilibrium.  Furthermore, novel phases in cold atom systems are conveniently
investigated by dynamical probes pushing the system out of equilibrium.
Here, we discuss the pumping of doubly-occupied sites in a fermionic Mott
insulator by a periodic modulation of the hopping amplitude. We show that
deep in the insulating phase the many-body system can be mapped onto an
effective two-level system which performs coherent Rabi oscillations due
to the driving.  Coupling of the two-level system to the remaining degrees
of freedom renders the Rabi oscillations damped. We compare this scheme
to an alternative description where the particles are incoherently pumped
into a broad continuum.
\end{abstract}

\pacs{03.75.Ss, 71.10.Fd, 31.15.aq}

\maketitle

\begin{figure}[t]
\includegraphics{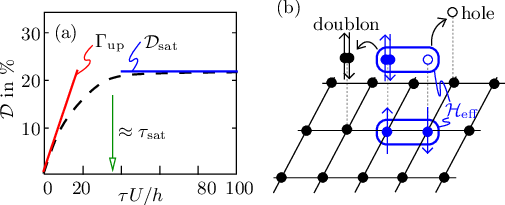}
\caption{%
(color online) (a) Sketch of an experimental sequence for the
generation of double occupancy by modulating the optical lattice
during a time $\tau$. After an initial buildup of doubly-occupied sites
with a rate $\Gamma_\text{up}$, the fraction of particles residing in
doubly-occupied sites saturates at a value $\mathcal D_\text{sat}$ after
a time $\tau_\text{sat}$. (b) Illustration of the effective Hilbert space
$\mathcal H_\text{eff}$ consisting of one bond embedded in an array of
sites. The bath degrees of freedom consist of all doublon-hole configurations
in which the pair resides on a different bond than where it was created.
}
\label{fig:experiment}
\end{figure}

Cold atoms in optical lattices provided us with numerous insights into the
paradigmatic quantum phase transition displayed by interacting lattice bosons
\cite{fisher:89,greiner:02}. Progress in the same setup with fermionic atoms
has now led to new, equally-exciting results in the study of the Hubbard
model \cite{scarola:08,leo:08,helmes:08,huber:09,kollath:06}. A decrease
of the compressibility when crossing over from the weakly to the strongly
interacting region was reported \cite{jordens:08,schneider:08}. Furthermore,
in the experiment by J\"ordens {\em et al.} \cite{jordens:08}, the strongly
correlated regime of fermions in an optical lattice has been investigated
by harmonically modulating the strength of the light beams forming the
optical lattice.  The modulation of the lattice generates doubly-occupied
sites (doublons). This generation of doublons deep in the Mott insulator,
where the interaction $U$ dominates over the hopping $t_{0}$, is the
focus of our present work. Experimentally, the fraction of particles on
doubly-occupied sites (or double occupancy $\mathcal{D}$) was measured
after modulating the optical lattice. For $U/6t_0=13.6$, a {\em saturation}
was observed at around $\mathcal{D}_\text{sat}=30\%$ after a modulation
time $\tau_\text{mod} \approx 50 h /U$ (50 ``cycles'') \cite{jordens:08}
which sets an upper bound on the saturation time $\tau_\text{sat}\leq
\tau_\text{mod}$ [see\ Fig.~\ref{fig:experiment}(a)].

We consider fermionic atoms, such as $^{40}$K, subject to an optical
lattice in the framework of the Hubbard model
\begin{equation}
H=-t_{0}\sum_{\langle i,j\rangle,\sigma}
\bigl(
c_{i\sigma}^{\dag}c_{j\sigma}^\pdag + {\rm H.c.}
\bigr)
+ U\sum_{i}
c_{i\uparrow}^{\dag}c_{i\uparrow}^\pdag
c_{i\downarrow}^{\dag}c_{i\downarrow}^\pdag,
\end{equation}
at half filling and with the interaction $U$ dominating over the hopping
amplitude $t_{0}$.  Here, the fermionic operators $c_{i\sigma}^{\dag}$ create
an atom in one of two possible hyperfine states $\sigma=\uparrow,\downarrow$
at site $i$ and the sum $\langle i,j\rangle$ runs over nearest neighbors.
The ground state of the system consists of one particle per site with
the hopping $t_0$ suppressed due to the strong on-site repulsion $U$. As
the temperature $T$ in the experiment is larger than the super-exchange
coupling $J=4t_0^2/U$ \cite{jordens:08}, no significant quantum correlations
over the bonds are present and the ground state of the system is largely
composed of singly-occupied sites ($T\ll U$) with a random spin-direction
on each site. A harmonic modulation of the lattice with frequency $\omega
\approx U/\hbar$ transfers energy to the system and (possibly) promotes
a particle across the bond, thereby creating a doubly-occupied site next
to an empty site, a doublon-hole pair.

\begin{figure}[t] 
\includegraphics{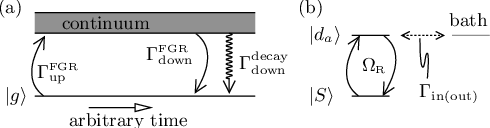}
\caption{%
Comparison of two possible descriptions: (a) A rate equation description
where an {\em incoherent} pump rate [$\Gamma_\text{up}^{\ssm FGR}$] into a
broad doublon-hole continuum and subsequent decay [$\Gamma_\text{down}^{\ssm
decay}$] or incoherent stimulated emission [$\Gamma_\text{up}^{\ssm FGR}$]
lead to a saturation. (b) {\em Coherent} Rabi oscillations [with frequency
$\Omega_{\ssm R}$] between the ground state and a single exited state where
decay out of the effective Hilbert space is introduced via coupling to a
bath [$\Gamma_\text{in(out)}$].
}
\label{fig:comparison} 
\end{figure}

Despite the fact that the Hubbard model represents a many-body problem
in a huge Hilbert space, we adopt a simple description via modeling a
single (typical) bond [Fig.~\ref{fig:experiment}(b)], and taking the
remaining bonds into account as an effective bath. Two different ideas
lend themselves to modeling the process of pumping doublon-hole pairs
[see Fig.~\ref{fig:comparison}]: (a) the direct excitation of the bond via
\emph{incoherent} transitions into the doublon-hole continuum with a rate
given by the Fermi golden rule
\begin{equation}
\label{eq:fgr}
  \Gamma^{\ssm FGR}_\text{mod} =  \frac{2\pi}{\hbar} 
  |\langle f | H_\text{mod} | g
  \rangle |^2 \rho(E_g+\hbar \omega);
\end{equation}
here, $E_g$ denotes the energy of the ground state, $\rho$ is the density of
states of the doublon-hole continuum, and $\langle f| H_\text{mod}|g\rangle$
is the matrix element arising from the modulation of the lattice. Taking the
modulation strength to be given by $\delta t$ and estimating the density
of states as $\rho \propto 1/t_0$, we obtain the scaling of the pump rate
$\Gamma^{\ssm FGR}_\text{mod} \propto \delta t^2/t_0$ as a function of
the hopping amplitude $t_0$ and the strength of the modulation $\delta
t$. (b) \emph{Coherent} Rabi oscillations in the isolated bond between
the ground state of the bond and the doubly-occupied excited state with
a Rabi frequency $\Omega_{\ssm R}\propto \delta t$ proportional to the
modulation strength. The Rabi oscillations generate doublon-hole pairs
and lead to a periodic modulation (with frequency $\Omega_{\ssm R}$) of
the double occupancy. Note that the rates of the two (extreme) processes,
incoherent pumping into a band (a) (with rate $\Gamma_\text{mod}^{\ssm
FGR}$) and coherent excitation of doublon-hole pairs (b) (with rate $\hbar
\Omega_{\ssm R}$), have a different dependence on the modulation strength
$\delta t$ and can therefore be discriminated experimentally.

Here, we propose a third way where the coherent Rabi oscillations produce a
doublon-hole pair and, subsequently, hopping of the doublon (or the hole)
to the neighboring sites leads to an incoherent decay of the excitation
into the doublon-hole continuum. The process is governed by two time scales,
the Rabi frequency $\Omega_{\ssm R} \propto \delta t$ which periodically
modulates the double occupancy and the decay  $\Gamma_\text{hop}\propto t_0$
which leads to a saturation.

A similar system, where a two-level atom is coupled simultaneously to a laser
field and to vacuum modes of the radiation field, is well-studied in quantum
optics \cite{tannoudji:77}. The two-level atom corresponds to the two
states on the bond and the laser plays the role of the harmonic modulation
of the lattice. However, in the atomic example, the spontaneously emitted
photon takes the atom back into the ground state where it continues with the
Rabi cycles, while in the present case the doublon (or hole) itself {\em
exits} the system. The present process then resembles more strongly that
of optical pumping into a third nondecaying level, with the accumulation
in the pumped level playing the role of the buildup in double occupancy.

%
%:fig: Comparison 

In the following, we first describe the effective model in the single bond
Hilbert space driven by a harmonic modulation of the lattice and derive the
frequency of the Rabi oscillations. We then account for the bath consisting
of the surrounding sites via a master equation approach and analyze it
in different limits.  We compare this approach to the situation where the
bond is directly excited into the doublon-hole continuum; see (a) above. It
turns out that the time scale ($\tau_{\ssm R}$) for the Rabi cycle is in
rough agreement with the experimental value of the saturation time, whereas
the treatment via a direct excitation into the doublon-hole continuum
leads to a time scale ($\tau_{\ssm FGR}$) which is an order of magnitude
larger ($\tau_{\ssm sat}^{\ssm exp}\lesssim \tau_{\ssm R} \ll \tau_{\ssm
FGR}$). For a quantitative agreement, one has to go beyond a single bond
in the description of the pumping process. Different bonds are coupled via
the pumping field, leading to a Dicke-like description \cite{dicke:54}, but,
most likely, many-body effects dominate the coupling of different bonds;
a systematic study of larger clusters deserves future efforts.

In deriving the effective Hamiltonian governing the dynamics in the Hilbert
space of a single bond (the sites are labeled by L and R), we neglect
the confining potential and focus on a homogeneous three dimensional
setup.  The ground state at half-filling involves one particle per site
without any spin order, i.e., the particles on the bond occupy the three
triplet states $ |T^{\pm 1}\rangle= c_{\ssmL\uparrow(\downarrow)}^{\dag}
c_{\ssmR\uparrow(\downarrow)}^{\dag}|0\rangle,\quad |T^{0}\rangle=
\frac{1}{\sqrt{2}} \bigl( c_{\ssmL\uparrow}^{\dag}c_{\ssmR\downarrow}^{\dag}+
c_{\ssmL\uparrow}^{\dag}c_{\ssmR\downarrow}^{\dag} \bigr) |0\rangle,
$ and the singlet state $ |S\,\rangle= \frac{1}{\sqrt{2}}
\bigl( c_{\ssmL\uparrow}^{\dag}c_{\ssmR\downarrow}^{\dag}-
c_{\ssmL\uparrow}^{\dag}c_{\ssmR\downarrow}^{\dag} \bigr) |0\rangle.  $
These states constitute the four degenerate ground states of a single
bond and are realized with equal probability of $1/4$ each. A harmonic
modulation of the lattice beam with $V=V_{0}+\delta V\cos(\omega\tau)$
leads to a time-dependent hopping Hamiltonian
\begin{equation}
  \label{eqn:hopp}
  H_\text{hop}= \delta t \cos(\omega\tau)
  \sum_\sigma
  \bigl(
  c_{\ssmL\sigma}^{\dag}c_{\ssmR\sigma}^{\pdag}+{\rm H.c.}
  \bigr),
\end{equation}
with $\delta t/t_{0}=\bigl(3/4-\sqrt{V_{0}/E_{r}}\bigr)\delta V/V_{0}$ and
$\tau$ denoting the time. This expression is valid for $\delta V/V_{0}\ll
1$ and $\delta V/V_{0} \ll \sqrt{E_{r}/V_{0}}$ \cite{hopping}. For
current experiments with fermions \cite{jordens:08}, where $\delta V/V_{0}
= 0.1$ and $V_{0}$ does not exceed $10 E_{r}$, both conditions are safely
fulfilled. In the derivation of Eq.~(\ref{eqn:hopp}), we have dropped the
constant hopping amplitude proportional to $t_0$, as hopping is blocked in
the Mott insulator at half-filling.  The hopping amplitude $t_0$ will be
reintroduced later as it is responsible for the decay of the doublon-hole
pair by coupling the bond to the surrounding lattice.  The modulation couples
the ground states to the doublon-hole states $ |d_{\ssm L(R)} \rangle =
c^\dagger_{{\ssm L(R)}\uparrow} c^\dagger_{{\ssm L(R)}\downarrow} |0\rangle,
$ with the doublon located on the left (right) site. These states exhibit
an energy offset $U$ with respect to the energy of the ground states. The
total Hamiltonian in the Hilbert space spanned by the four ground states
(triplet and singlet) and the two doublon-hole pairs is given by
\begin{align}
  H_\text{eff}&=
  \sum_{\alpha={\ssm L, R}} U |d_{\alpha}\rangle\langle d_{\alpha} | 
  + H_\text{hop} \approx \sum_{\alpha={\ssm L, R}}
U|d_{\alpha}\rangle\langle d_{\alpha} |
\nonumber
\\
\label{eq:ham}
&\qquad \qquad+
\delta t 
\Bigl[
e^{i\omega\tau/\hbar} | d_\text{a}\rangle\langle S| + 
e^{-i\omega\tau/\hbar} |S\rangle\langle d_\text{a}|
\Bigr] ,
\end{align}
where the first term describes the energy offset by $U$ and the second
term the driving due to the lattice modulation (\ref{eqn:hopp}). In
going from the first to the second equation in (\ref{eq:ham}), the
rotating wave approximation has been used and we have introduced
the {\em active} and {\em passive} doublon-hole states given by
$|d_\text{a(p)}\rangle=(|d_{\ssm L}\rangle \pm |d_{\ssm R}\rangle)/\sqrt{2}$;
these are related to the left (right) doublon-hole state via a basis change.
Driving the system on resonance with $\hbar \omega=U$, the singlet state
is coupled to the excited state via Rabi oscillations. Within the present
description, off-resonant coupling can be described in the standard
manner. As can be seen from Eq.~(\ref{eq:ham}), the triplet states are
unaffected by the lattice modulations.  Only the singlet state is coupled
to the active doublon-hole pairs $|d_\text{a}\rangle$. Going over to the
interaction representation with the reference dynamics given by the first
term in (\ref{eq:ham}), the relevant Hamiltonian in the effective two-level
system $\mathcal{H}_\text{eff}= \text{span}\{ |d_\text{a}\rangle,|S\rangle\}$
can be expressed as a matrix
\begin{equation}
\label{eqn:hammatrix}
\mathsf H = 
\begin{pmatrix}
0 & \delta t \\
\delta t & 0
\end{pmatrix},
\end{equation}
which describes the coherent oscillations in the two-level system with
the Rabi frequency
\begin{equation}
\Omega_{\ssm R}=\frac{2\delta t}{\hbar}.
\end{equation}
The triplet states and the passive doublon-hole pair remain eigenstates
of the time-dependent Hamiltonian and therefore particles in such states
are not driven.

So far, we have neglected the effect of $t_{0}$ on the excited state
$|d_\text{a}\rangle$. Provided the bond is in a doublon-hole state, we
can lose the state from our single bond description via the hopping of the
doublon (or the hole) out of the bond; cf.\ Fig.~\ref{fig:experiment}(b)
\cite{vacuum}.  This effect is captured by a bath representing the states
with the doublon and hole separated from each other. Note that the bath
states still contribute to the double occupancy (as there is still a
doublon present) but do not participate in Rabi oscillations (as the
doublon and the hole are not located on the same bond). We denote the
rate at which the doublon-hole bond decays into a separated doublon and a
hole by $\Gamma_\text{out}$, and the reverse rate at which a doublon and
a hole collide to form a doublon-hole pair by $\Gamma_\text{in}$. Taking
the coupling to the bath into account and performing a standard analysis
similar to \cite{gardiner:5} leads to an incoherent term (besides
the standard term proportional to $[\mathsf{H},\rho]$) in the evolution
of the reduced density matrix $\rho$ in the effective Hilbert space
$\mathcal{H}_\text{eff}= \text{span}\{ |d_\text{a}\rangle,|S\rangle\}$
of the form
\begin{equation}
	\label{eq:incoh}
\frac{\partial\rho}{\partial\tau}
\bigg|_\text{inc}
=
-\frac{1}{2} 
\begin{pmatrix}
  2(\Gamma_\text{out} \rho_{11} - \Gamma_\text{in} \rho_0)
  & \Gamma_\text{out} \rho_{12} \\
 \Gamma_\text{out}\rho_{21} & 0
\end{pmatrix}.
\end{equation}
Different from a standard master equation \cite{gardiner:5},
Eq.~(\ref{eq:incoh}) is not probability preserving, since with
probability $\rho_0 = 1-\rho_{11} -\rho_{22}$ the doublon and the
hole have separated and thus have left the effective Hilbert space
$\mathcal{H}_\text{eff}$. In the steady state situation with $\partial_\tau
\rho=0$, the probability that the system contributes to the double occupancy
is given by
\begin{equation}
	\label{eq:sat}
  \rho_{\ssm D}^\text{sat}= \rho_0^\text{sat} + \rho_{11}^\text{sat} =
  1-\rho^\text{sat}_{22} = \frac{\Gamma_\text{out}+ \Gamma_\text{in}} 
  {\Gamma_\text{out} + 2 \Gamma_\text{in}},
\end{equation}
the sum of the probabilities that the system is in the doublon-hole
pair state ($\rho_{11}$) or in the bath ($\rho_0$). Note that atoms
which left the effective Hilbert space still contribute to the double
occupancy. The bath can be pictured as a gas of doublons and holes (with
equal numbers). Initially, the bath does not contain any particles such
that $\Gamma_\text{in}=0$ (this approximation remains valid as long as
the density of particle is low); furthermore, we estimate the escape
rate by $\Gamma_\text{out}=\Gamma_\text{hop} \sim t_{0}/\hbar$. For
$\Gamma_\text{in}=0$, Eq.~(\ref{eq:sat}) provides the saturation value
$\rho_{\ssm D}^\text{sat}=1$, i.e., the singlet is certainly transferred
to one of the doublon-hole states.
%
%:fig: Main result
\begin{figure}[t]
\includegraphics{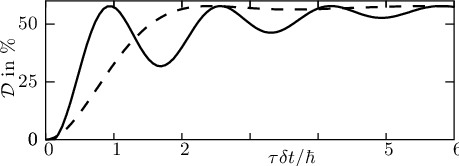}
\caption{%
Fraction of particles residing in doubly-occupied sites after modulation time
$\tau$. After a buildup time corresponding to approximately one $\pi$-pulse
[$\tau_{\ssm R}=\pi/\Omega_{\ssm R}=\hbar\pi/2\delta t$], the decay,
with (weak) strength $\Gamma_\text{out} = \delta t/\hbar \approx 0.24
t_0/\hbar$ and $\Gamma_\text{in}=0$, leads to a saturation at $\mathcal
D_\text{sat}=58\%$ corresponding to the fact that all ``active'' bonds,
singlets, are excited in the system (solid line). A more realistic situation
where the damping is increased by a factor of 2 renders the oscillations
barely visible (dashed line). Taking into account a finite value of
$\Gamma_\text{in}$ leads to a decrease of the saturation value. Note that,
irrespective of the saturation value, the time scale needed to pump the
system into a doubly occupied state is given by the Rabi frequency.
}
\label{fig:solutions}
\end{figure}
Using the dilute gas approximation, $\Gamma_\text{in}=0$, we can solve
the equation of motion for the density matrix
\begin{equation}\label{eq:time_evol}
  \partial_{\tau}\rho=\frac{1}{i\hbar}[\mathsf H, \rho]-
\frac{\Gamma_\text{out}}{2} 
\begin{pmatrix}
  2 \rho_{11} & \rho_{12} \\
\rho_{21} & 0
\end{pmatrix}.
\end{equation}
Assuming that initially the bond is in a singlet state
\begin{equation}
\rho(\tau=0)=
\begin{pmatrix}
0 & 0 \\
0 & 1
\end{pmatrix},
\end{equation}
Eq.~(\ref{eq:time_evol}) gives the time evolution of the density matrix
of the bond. Qualitatively, the weight is transferred to the exited
state $|d_\text{a}\rangle$ at a rate $\hbar \Omega_{\ssm R}$ and then
the decay into other states occurs with a rate $\Gamma_\text{out}$ (cf.\
Fig.~\ref{fig:solutions}).

To calculate the fraction of particles in a doublon state, we need to know
how many bonds are initially in a singlet configuration. We estimate the
probability that a specific site is part of a singlet as $P_S = 1- (3/4)^{d}
\approx 0.58 $, here $d=3$ is the ``dimension'' of the lattice modulation
(it counts the number of bonds per site) and we have assumed that the
singlets are independent of each other and occupied with a probability
$1/4$. The fraction of particles in a doublon state is given by
\begin{equation}\label{eq:doublon}
  \mathcal{D}=P_S \rho_{\ssm D}(\tau)
\end{equation}
with $\rho_{\ssm D}(\tau)=1-\rho_{22}(\tau)$ the probability that
this singlet is transferred to a doublon state. The system quickly
reaches a steady state with $\mathcal{D}_\text{sat}=P_S$ (cf.\
Fig.~\ref{fig:solutions}).  In three dimensions, we reach a maximal value
$\mathcal D_\text{sat}=58\%$. Note that at this value of double occupancy the
dilute gas approximation $\Gamma_\text{in}=0$ becomes invalid. We expect that
$\Gamma_\text{in}$ increases for increasing double occupancy and, therefore,
the double occupancy saturates at a lower value. A theory that aims at a
quantitative result for the saturation value should take such effects into
account and solve the system self-consistently. Assuming that for long times
the density of particles in the bath is high such that $\Gamma_\text{in}
\gg \Gamma_\text{out}$, we obtain the result that $\rho_{\ssm D}^\text{sat}
= 50\%$ [cf.\ Eq.~(\ref{eq:sat})] and $\mathcal{D}_\text{sat}=29\%$,
comparable to experimental results \cite{jordens:08}. The time scale needed
to excite the system is approximately given by the time $\tau_{\ssm R}
= \pi/\Omega_{\ssm R}$ of a $\pi$ pulse, irrespective of the saturation
value obtained. Note that the finite line-width of the perturbing laser,
the scattering between doublons, and an inhomogeneous environment for a
single bond due to the trap are other processes which lead to a dephasing of
the coherent single bond dynamics. As the effect of $t_{0}$ on the excited
state leads already to an over-damped dynamics, these other effects are
neglected here.

%:Comparison to a rate equation approach 
%
Another approach to describe the saturation in the double occupancy is
based on a rate equation for the doublon fraction. The {\em rates} at which
excitations are generated are taken into account via a rate equation for
the doublon concentration
\begin{equation}
\label{eqn:rateequation}
\partial_{\tau} {\mathcal D} = 
(1-{\mathcal D})\Gamma_\text{up}
-{\mathcal D} \Gamma_\text{down}.
\end{equation}
Within this description two independent processes are assumed. First,
the system is excited with a rate $\Gamma_\text{up}$. Second,
at a later uncorrelated stage, these excitations are removed with
a rate $\Gamma_\text{down}$ [cf.\ Fig.~\ref{fig:comparison}(a)].
In order to compare the two schemes, the rate equation and the Rabi
oscillation, we extract the time scales of both processes. The Fermi
golden rule matrix element is approximately given by $\delta t^{2}$
\cite{huber:09}. Additionally, we need an estimate for the density of
states. Once excited, the doublon as well as the hole can move freely
through the lattice. The corresponding doublon-hole continuum acquires a
width of approximately $24t_{0}$ (in three dimensions). A crude estimate
for the density of states is then given by $\rho\approx1/24t_{0}$. We
then arrive at the two time scales $\tau_{\ssm R} = \pi/\Omega_{\ssm
R}=\pi\hbar/2\delta t$ and $\tau_{\ssm FGR} = (\hbar/2 \pi) 24 t_{0}/\delta
t^{2}= (24 t_0/\pi^{2}\delta t)\tau_{\ssm R} \approx 10 \tau_{\ssm R}$,
associated with a Rabi oscillation and the rate equation, respectively. For
the last step, we have used $\delta t/t_{0}\approx 0.24$ corresponding to
current experiments. Note that the time scale to reach saturation in the
incoherent Fermi golden rule approach is an order of magnitude larger than
the one predicted by the Rabi oscillation picture. In Ref.~\cite{jordens:08}
the saturation sets in before 50 ``cycles'', which is even smaller than
$\tau_{\ssm R}$ [$\tau_\text{mod} = 50 h/U = (200 \delta t/U)\tau_{\ssm R}
\approx 0.6 \tau_{\ssm R}$ for $U/6t_0=13.6$ and $\delta t/t_{0}\approx
0.24$], promoting the picture of the coherent Rabi oscillations. In an
experiment, one can test one theory against the other by changing the
coupling strength $\delta t$ while keeping all other parameters, such
as the bandwidth $1/24t_0$, constant. The time scale for the saturation
scales as $\delta t^{-2}$ in the Fermi golden rule picture, whereas Rabi
oscillations produce a saturation time inversely proportional to $\delta t$.

In summary, we have discussed two alternative explanations for the buildup
and the saturation of the double occupancy in the modulation experiments of
Ref.~\cite{jordens:08} deep in the Mott insulating phase. The comparison to
experiment favors the description in terms of an effective two-level
system with coherent driving, leading to the characteristic time scale of
$\tau_{\ssm R}\approx \pi/\Omega_{\ssm R}=\hbar\pi/2\delta t$ for half a Rabi
cycle. Given the large decay rate proportional to $t_{0}>\hbar \Omega_{\ssm
R} \approx \delta t$, only one $\pi$-pulse is observable. The predictions can
be experimentally tested by studying the dependence of the saturation time on
various system parameters as $V_{0}/E_{r}$, $\delta V/V_{0}$, and $t_{0}/U$.

We acknowledge fruitful discussions with E.\ Altman, G.\ Blatter, H.~P.\
B\"uchler, the group of T.\ Esslinger, S.\ F\"olling, L.\ Goren, and A.\
R\"uegg. This work was supported by the Swiss National Foundation through
the NCCR MaNEP.


\begin{thebibliography}{10}
\bibitem{hopping}
When deriving (\ref{eqn:hopp}), care has to be taken in expanding the
  exponential part of the hopping amplitude \cite{zwerger:03}.
\bibitem{vacuum}
We think about the singly-occupied lattice as being the vacuum with excitations
  given by doublons and holes.
\end{thebibliography}
\end{document}